\documentclass[sigconf,authorversion,nonacm]{acmart}
\AtBeginDocument{%
  \providecommand\BibTeX{{%
    \normalfont B\kern-0.5em{\scshape i\kern-0.25em b}\kern-0.8em\TeX}}}

\begin{document}

\title{Combining Cloud and Mobile Computing for Machine Learning}


\author{Ruiqi Xu}
\email{ruiqix@uchicago.edu}
\affiliation{%
  \institution{University of Chicago}
  \city{Chicago}
  \state{Illinois}
  \country{United States}}

\author{Tianchi Zhang}
\email{tonyztc@uchicago.edu}
\affiliation{%
  \institution{University of Chicago}
  \city{Chicago}
  \state{Illinois}
  \country{United States}}


\begin{abstract}
  Although the computing power of mobile devices is increasing, machine learning models are also growing in size. This trend creates problems for mobile devices due to limitations like their memory capacity and battery life. While many services, like ChatGPT and Midjourney, run all the inferences in the cloud, we believe a flexible and fine-grained task distribution is more desirable. In this work, we consider model segmentation as a solution to improving the user experience, dividing the computation between mobile devices and the cloud in a way that offloads the compute-heavy portion of the model while minimizing the data transfer required. We show that the division not only reduces the wait time for users but can also be fine-tuned to optimize the workloads of the cloud. To achieve that, we design a scheduler that collects information about network quality, client device capability, and job requirements, making decisions to achieve consistent performance across a range of devices while reducing the work the cloud needs to perform.
  
\end{abstract}




\maketitle

\section{Motivation}

    While the successes of services based on large machine learning models like ChatGPT, Stable Diffusion, Midjourney, and DALL-E draw attention to the power of AI, a constant trend in the field is to make the models bigger. A survey found that the language models have grown by seven orders of magnitude from 1950 to 2018, and from 2018 to 2022, they have increased by another five orders of magnitude. Although vision models grow at a constant rate, the size still expands by seven times in this period\cite{RN2}. Despite the fact that this trend makes running those models on mobile devices seem impossible, we cannot ignore the advancement of mobile accelerators. The developers of the MLPerf mobile inference benchmark found a 2x throughput increase and 12x latency reduction on mobile devices in a period of 6 months\cite{RN1}. While many companies choose to host their services entirely on the cloud, we believe this decision gives up many optimization opportunities. According to a recent work studying network latency using Google's Stadia cloud-gaming service, the round-trip time (RTT) values are consistently lower than 25 ms\cite{RN3}. At the same time, the 5G wireless network technology is aiming for end-to-end latency on the order of 1 ms\cite{5G}. This means users can exchange data with the cloud at little cost. Thus, we think it is possible to distribute the workload in a more flexible fashion by sharing the results of layers in the middle of a model, essentially allowing the cloud to collaborate with mobile devices on inference tasks.

\section{Introduction}
    
    The idea behind our work is that by sharing intermediate results of layers in a neural network, we allow a different device to pick up the leftover work, effectively dividing the inference task. We aim to study the effectiveness of model segmentation using two types of tasks. The first one is image recognition. To achieve high accuracy on classification tasks like ImageNet, many models require a large number of parameters. These models are usually formed by a set of independent residual layer blocks, each taking in the output of its predecessor. This structure makes it easy to grab the output of one residual layer block and share it with another device. However, most of these models already run well on mobile devices with machine learning accelerators, and many image recognition tasks are latency-sensitive. Therefore, we need to make good decisions on the model and offloaded layers to obtain speedup. The second type we target is generative models. In particular, we pick stable diffusion because it is not only popular but also open-source and readily available. These models are usually larger and more compute-intensive, giving us additional room for various design choices.
    
    After getting a clear idea of the cost, we will build a system that allows us to send the intermediate results of an inference step and finish the job on a different machine. We want to ensure that doing so does not damage the quality of the output. Based on the validation result, we can augment the design to pass the results over the internet, evaluating the network latency and overall performance. These observations should enable us to model the cost and benefits of various decisions and formulate an algorithm for task assignment. For devices with slower speeds or poor network connectivity, we will put more work into the cloud, and vice versa. The goal is to perform as little work on the cloud as possible while satisfying a service-level agreement that defines a maximum end-to-end latency.

    The solution to this question would unlock some interesting possibilities. First, due to the improving performance of mobile devices, we believe that for certain models, we can dynamically decide the location of the computation, putting some or all of the computation on mobile devices without significant loss in quality or latency. Apart from alleviating the burden on the cloud, it transforms the nature of these workloads from "production" to "non-production" jobs, as described by Google\cite{RN4}, enabling more aggressive over-subscription by cloud providers. When there are higher-priority jobs, we reduce the portion of cloud-accelerated work while ensuring that all the users get some quality guarantees. Second, for different mobile devices, we can alter the amount of computation performed on the server based on mobile devices' processing power to distribute cloud resources more evenly, producing consistent user experiences. Third, we hope to enable the cloud to adapt to the rapidly changing landscape of mobile computing, dynamically altering resource allocations to minimize waste.

\section{Background}
\subsection{Large Machine Learning Models}
    The size of machine learning models has kept increasing before the term ``large machine learning models'' ever existed. One milestone of this process is the residual network (ResNet)\cite{resnet}, which showed the ability to address the vanishing gradient problem. The ResNet is an image classification model, and it can achieve higher ImageNet validation accuracy by stacking more residual building blocks, which inspires other models. The idea of the residual block is widely used in newer models like ResNeXt\cite{resnext}, BERT\cite{BERT}, and AlphaGO\cite{alphago}, which achieve great performance in areas including computer vision, natural language processing, and gaming. It is also used in both models, RegNet and Stable Diffusion, we are experimenting, especially in RegNet\cite{regnet1, regnet2}. RegNet is a successor of ResNet, created by applying neural architecture search on how to build residual blocks and how to connect residual blocks to create a model with size limitations.

    With the size of models increasing, we notice that we can go beyond the traditional classification task and design and train machine learning models for more complex tasks, like drug discovery\cite{drug1, drug2} and self-driving\cite{car}. One of the most popular tasks is the generative task. Current generative machine learning models have impressive performance, especially those with transformer architectures. Some of them are earning significant profits, like ChatGPTs\cite{GPT}, DALL·E\cite{DALLE}, and Stable Diffusion. Stable Diffusion\cite{stablediff} is one of the models we will experiment with. It is released by stability.ai and can generate detailed images based on text descriptions.

\subsection{Machine Learning Acceleration}

    The rapid growth of services relying on machine learning has created a huge demand for specialized accelerators, ranging from professional GPUs used in data centers to energy-efficient ones for mobile devices. NVIDIA recently released their H200 series GPU. It claims a peak TF32 performance of 989 TFLOPS and FP16 performance of 1,979 TFLOPS. It is also equipped with 141GB of HBM3e for training large models\cite{H200}. On the mobile computing side, the neural engine on iPhone 15 Pro's A17 Pro processor has an advertised performance of 35 trillion operations per second\cite{A17}. On the Android side, Qualcomm's latest Snapdragon 8 Gen 3 SoC also boasts a 98\% performance improvement for its NPU compared to the previous generation\cite{8Gen3}.
    
\begin{figure}[h]
    \centering
    \includegraphics[width=8cm]{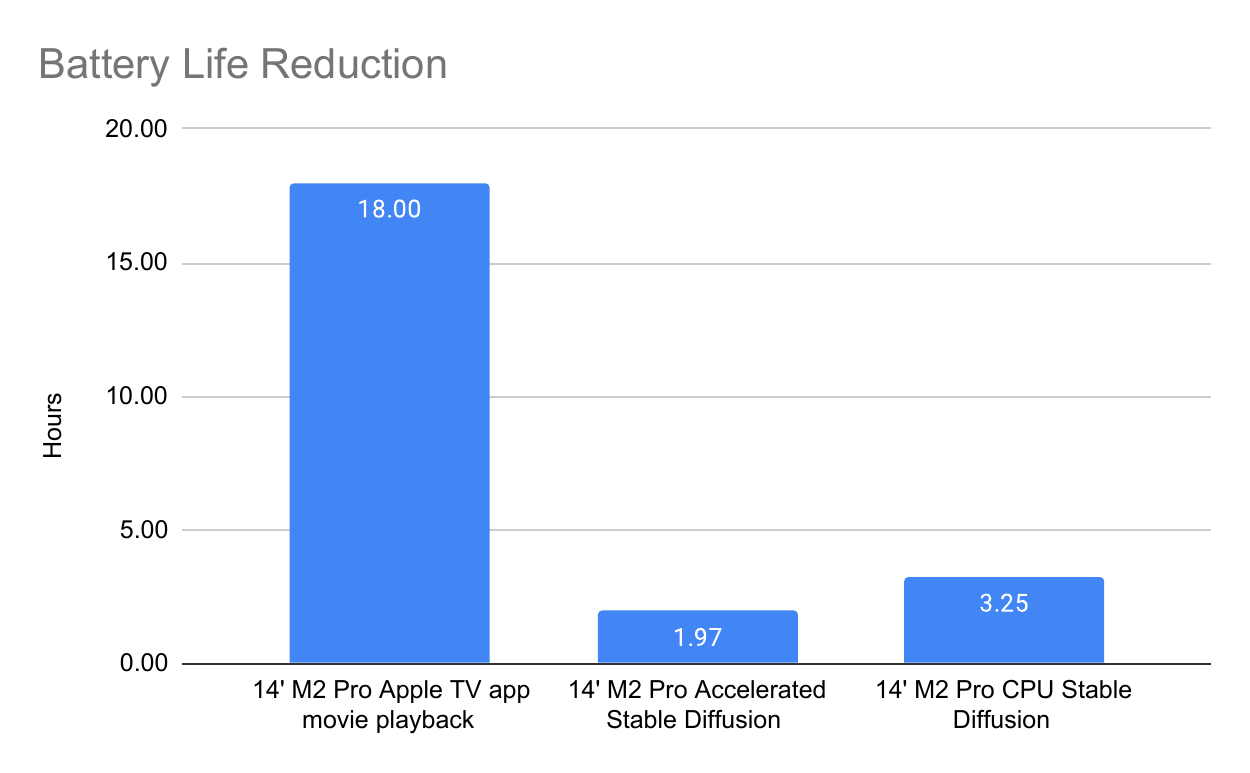}
    \caption{Battery Life Reduction of Running Large Machine Learning on Edge}
    \label{fig: battery}
\end{figure}

    While these numbers look as if mobile platforms can handle increasingly complex machine-learning tasks, we need to consider a few important factors. First, the size of video memory in most mobile GPUs does not keep up with the size of the models. Even a size-conscious model like stable diffusion requires more than 4GB of memory\cite{StableDiffusionWiki}. This means a large number of mobile GPUs and high-end desktop GPUs from a few years ago can no longer run it natively. Although Apple's unified memory system offers some relief, having an application take up so much RAM can have a detrimental effect on user experience. Second, most mobile devices have limited battery life. Running large amounts of computation can lead to overheating and high energy consumption. We run a Stable Diffusion model on a 14-inch Macbook Pro with M2 Pro and the result is presented in Figure \ref{fig: battery}. Compared with the video streaming time provided by Apple\cite{14mbp}, the battery life is reduced to 11\% or 18\% w/wo using the neural engine on M2 Pro. Third, even though many recent works are running quantized models on mobile devices, the performance of the models is more or less sacrificed\cite{hsquant, cnnquant}. Therefore, we believe that for latency-insensitive jobs, offloading to the cloud still offers significant advantages.

\subsection{Low-Latency Inference Services}
    Since machine learning inference services can be time-sensitive, many research works focus on reducing the latency of those tasks. TensorFlow-Serving is Google's solution to this problem\cite{TensorFlow-Serving}. It offers a library to allow users to manage multiple batching queues so that they run efficiently. However, its main focus is still offering an easy interface and tools like the version manager. The Volex data management system includes more optimizations, demonstrating the effectiveness of caching inference results as well as partitioning and replicating the tables in improving prediction latency\cite{Velox}. Clipper is a low-latency prediction serving system that aims to achieve better throughput and accuracy without modifying the machine learning models\cite{Clipper}. It also caches the model predictions for certain inputs to avoid computation. It makes batching decisions intelligently to amortize startup costs while staying within the SLA. Additionally, it allows predictions to be made without waiting for slower models using a straggler mitigation technique. The Nexus paper also proposes some interesting optimizations\cite{Nexus}. In addition to adopting the intelligent batching technique discussed above, it groups DNN invocations as queries so that they can be optimized for throughput. Furthermore, they allow partial batching of the network for better performance. Lastly, the work LASER takes advantage of the fact that logistic regression fails gracefully when features are missing\cite{LASER}. It replaces entries with their expected values when data accesses take too long to stay within the SLA.

\subsection{Machine Learning on the Cloud}

    Due to the computationally intensive nature of machine learning, past research has mainly focused on accelerating the training and inference using powerful servers. As early as 2012, researchers have started trying to use tens of thousands of CPU cores to train deep networks with billions of parameters, proposing algorithms for asynchronous stochastic gradient descent and distributed batch optimizations\cite{LargeScaleDistributed}. TensorFlow proposes the approach of representing computation, shared states, and operations on those states as a dataflow graph, allowing different portions of it to be mapped across machines in a cluster\cite{TensorFlow}. The improvement in speed and flexibility led to numerous innovative algorithms and optimizations. 

    While these innovations laid the groundwork for the success of machine learning, the challenges of bringing state-of-the-art models closer to the users remain. The developers of MAUI have shown that offloading tasks to the Cloud can significantly expand the capability of mobile devices\cite{MAUI}. There is also no shortage of investigations focusing on machine learning. The MDINFERENCE paper develops a system that selects different models based on network conditions to maximize accuracy without violating SLA\cite{MDINFERENCE}. The paper on Distributed Deep Neural Networks\cite{DDNN} aims to map different sections of DNN onto a series of devices, showing that it reduces the amount of data that needs to be shared. However, the models they consider are much smaller, meaning they only need to involve the cloud when confidence is low. We target more complex models, focusing more on the latency aspect. The Neurosurgeon work \cite{Neurosurgeon} examines how to dynamically partition DNNs based on network quality, server load, and mobile device performance. However, while they aim to lower latency, we try to meet the latency target while keeping server utilization low. Furthermore, we limit the granularity of partitioning so that the server does not need to handle diverse requests. To further reduce the workload on the edge, another paper shows that the portion of the work that remains on the edge can run a quantized version of the network, which requires less computational resources and memory\cite{CollaborativeInference}.

\section{Implementation}

\subsection{Model Splitting}

\subsubsection{RegNet}
\begin{figure}[h]
    \centering
    \includegraphics[width=8cm]{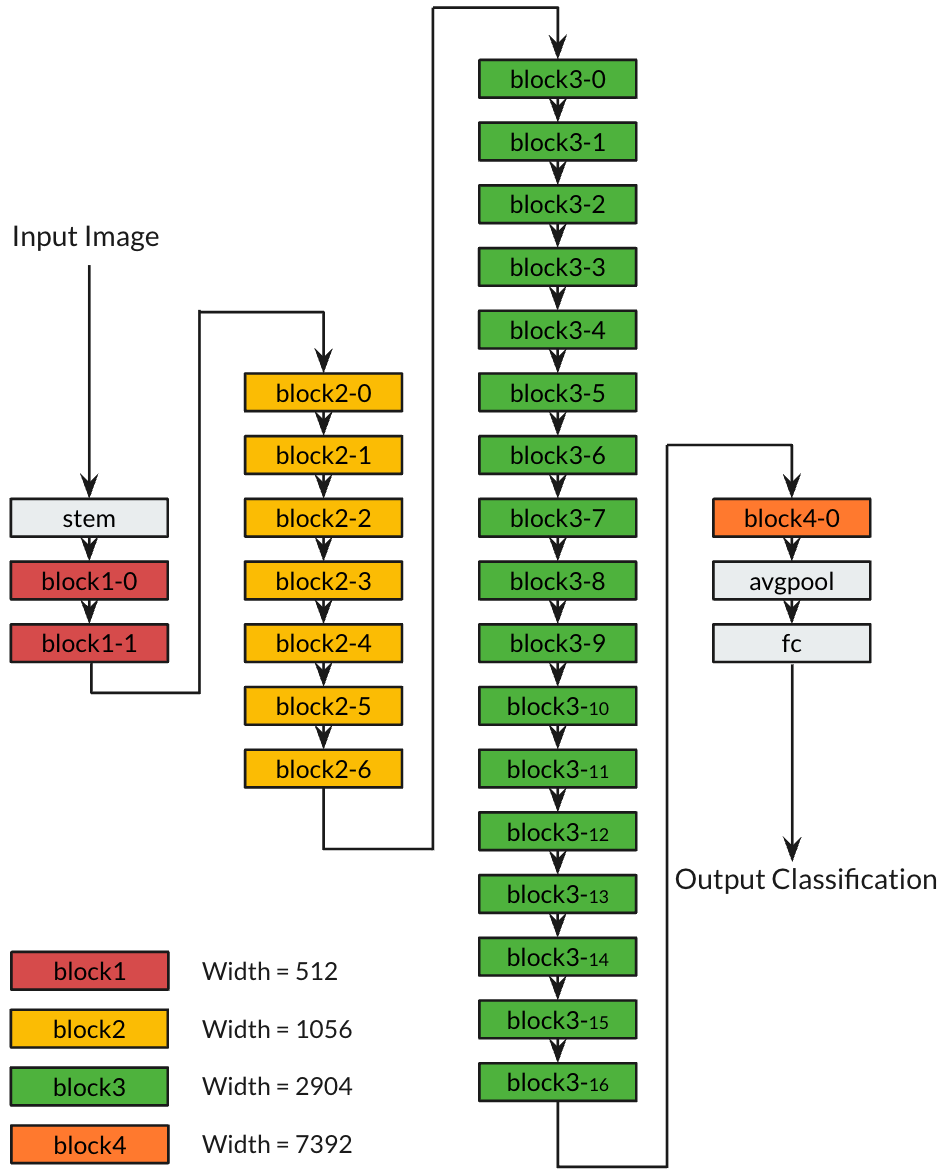}
    \caption{Model architecture of RegNet}
    \label{fig: regnet_arch}
\end{figure}
\begin{table}[h]
    \centering
    \begin{tabular}{c|c|c}
        \hline
        \hline
        Splitting Point & Tensor Shape & Tensor Size (kilobyte)\\
        \hline
        stem & 1x32x192x192 & 4608\\
        \hline
        block1 & 1x528x96x96 & 188496\\
        \hline
        block2 & 1x1056x48x48 & 9216\\
        \hline
        block3 & 1x2904x24x24 & 5202\\
        \hline
        block4 & 1x7392x12x12 & 41472\\
        \hline
        avgpool & 1x7392x1x1 & 29\\
        \hline
        \hline
    \end{tabular}
    \caption{Tensor size of splitting points in RegNet}
    \label{tab: regnet_arch}
\end{table}

    The model we picked comes from a recently published paper\cite{RegNet}. RegNet uses four different kinds of residual blocks, painted in different colors in figure \ref{fig: regnet_arch}. We insert splitting points after each kind of block is finished and name these splitting points by the block in front of it. The size of the tensors we need to transfer from cloud to edge devices is listed in the table \ref{tab: regnet_arch}.

\subsubsection{Stable Diffusion}
\begin{figure}[h]
    \centering
    \includegraphics[width=6cm]{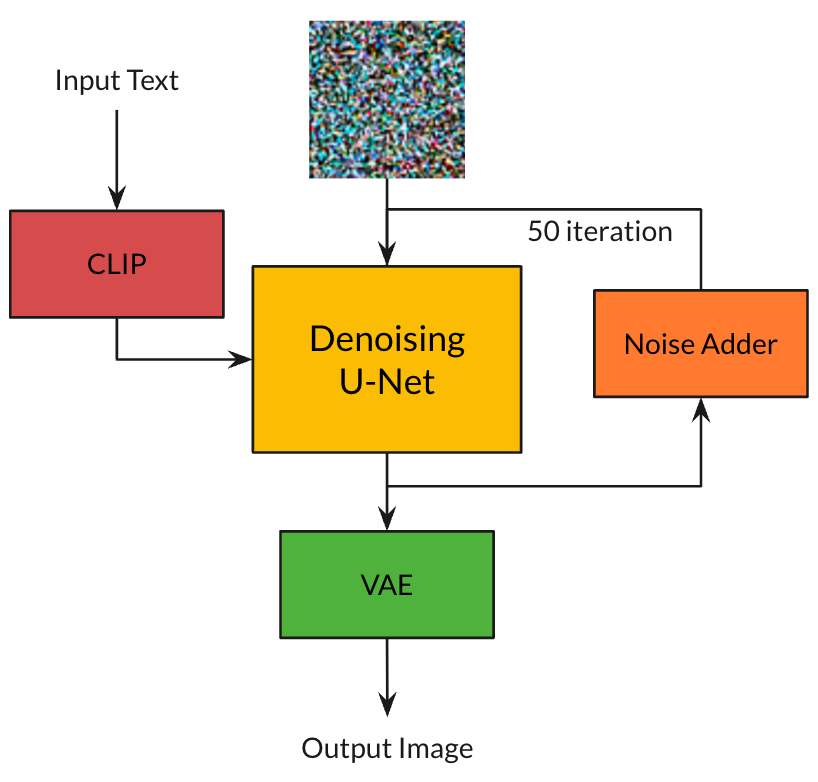}
    \caption{Model architecture of Stable Diffusion}
    \label{fig: stable_diff_arch}
\end{figure}
\begin{table}[h]
    \centering
    \begin{tabular}{c|c|c}
        \hline
        \hline
        Splitting Point & Tensor Shape & Tensor Size (kilobyte)\\
        \hline
        denoising0 & 2x77x768 & 232 \\
        \hline
        denoising5 & 4x64x64 & \\
        to & + & 296\\
        denoising45 & 2x77x768 & \\
        \hline
        denoising50 & 4x64x64 & 64\\
        \hline
        \hline
    \end{tabular}
    \caption{Tensor size of splitting points in Stable Diffusion}
    \label{tab: stable_diff_arch}
\end{table}

    The model we picked comes from a recently published paper\cite{stablediff}. The process of the stable diffusion model is presented in figure \ref{fig: stable_diff_arch}. It generates a block of Gaussian noise and uses CLIP\cite{clip} to generate the text embedding. It then runs the denoising U-Net for fifty iterations, with additional noise added after each iteration. It then uses a Variational Autoencoders (VAE) block to generate the output image. We insert splitting points after each five iterations of the denoising loop and also between the U-Net and the VAE block. These splitting points are named by the number of iterations completed. The size of the tensors we need to transfer is listed in the table \ref{tab: stable_diff_arch}.

\subsection{Data Transmission}

    To transmit the data between the server and the client, we chose the TCP protocol. Compared to UDP, TCP offers ordered, reliable, and error-checked delivery of information\cite{TCP}. The protocol is connection-oriented, meaning the connection must be established before any communication can happen. The handshaking, retransmission, and error detection can add to the latency, but they are critical for the correct functionalities. A custom implementation based on UDP might have better performance, but should not make significant differences. For the specific implementations, we use the Python socket and socketserver library. The socketserver library provides a handler class. When it receives a request, it automatically invokes the functions we defined, which will start the stable diffusion pipeline. The socket library is used by the client, allowing it to reconstruct the payload.

    Since these libraries do not handle complex data types, we need to serialize data structures like PyTorch tensors. Luckily, PyTorch has a save function, which dumps the content of a tensor as bytes into a Python BytesIO buffer. We have shown that the resulting data can be correctly reconstructed using PyTorch's load function. While the shape information might be lost, we assume the client is aware of the expected dimensions and recover that. We also include an evaluation of the costs of these steps in later sections.

\subsection{Enforcing End-to-end Latency}

    Based on the experiments we performed on the stable diffusion model, we can write the end-to-end latency as follows \[\frac{n_{cloud}}{r_{cloud}}+\frac{n_{total}-n_{cloud}}{r_{dev}} + t_{network} + t_{decode}\]

    Here $n_{cloud}$ is the number of diffusion iterations we want to run on the cloud, and $n_{total}$ is the total number of iterations necessary to get high-quality output. $r_{cloud}$ and $r_{dev}$ are the number of diffusion iterations per second each machine can perform. The $t_{network}$ is the network latency in the round-trip. $t_{decode}$ is the amount of time it takes to decode the embedding on the device. To simplify the problem, we assume that the decoding speed is proportional to the inference speed. Thus, we will rewrite $t_{decode}$ as $k_{decode}/r_{dev}$. Notice that we ignore the encoding cost because our experiments find it to be negligible.

    Then, we can rearrange the equation to derive $n_{cloud}$.

    \[n_{cloud} = \frac{r_{cloud}r_{dev}}{r_{dev}-r_{cloud}}(t_{lim}-\frac{k_{decode}}{r_{dev}} - t_{network}) -\frac{r_{cloud}}{r_{dev}-r_{cloud}}\]

    Based on that, we create a simple algorithm for determining how many iterations we run on the cloud. To group different requests together, we will increment the number of iterations that the cloud runs in fixed steps. For a request, we calculate $n_{cloud}$ for it, round up the value, and assign it in the following way:
    \[n_{final} = ceil(n_{cloud}) + (n_{step} - (n_{cloud} \% n_{step}))\]

    We will show in the next section of this paper that the simple design is quite effective for a set of requests from a range of devices.

    Note that in a more realistic setup, we also need to consider the queuing time, but since it is a constant at evaluation time, we can simply modify the equation as follows.
    \[\frac{n_{cloud}}{r_{cloud}}+\frac{n_{total}-n_{cloud}}{r_{dev}} + t_{network} + t_{decode} + t_{queue}\]
    To reduce the complexity of the problem, we do not consider queuing time in this paper.

\subsection{Intelligent Batching}

    Batching is an important technique for improving GPU utilization in machine learning tasks. The effect of batching can be explained using the simple equation below.
    \[t_{batch} = t_{startup} + t_{task}*n_{batch}\]
    Here $t_{task}$ is the increase in time consumption from each additional task and $n_{batch}$ is the batch size. Essentially, batching allows us to amortize some fixed costs in invoking the model over multiple examples, reducing the waste and increasing the effective throughput. However, one thing we need to consider is that batching also increases the response time of a task. If we put two jobs that can start at the same time on two GPUs into one batch and run them on one GPU, the processing latency for both tasks will increase. Thus, to make up for the increase in time consumption, we might have to run additional cycles on the Cloud, offsetting the increase in efficiency.

    Luckily, the model we proposed can capture this increase in cost effectively. 
    \[\frac{n_{cloud}}{r_{cloud} / c_{batch}}+\frac{n_{total}-n_{cloud}}{r_{dev}} + t_{network} + t_{decode}\]
    The added factor $c_{batch}$ is introduced to capture the decrease in GPU processing speed caused by batching. In the evaluation section, we will see how the stable diffusion model responds to batching. Furthermore, we can investigate how the number of jobs that can be batched together changes in response to various batching costs assuming that we do not increase the number of iterations that we perform for them. The reasoning behind this is that the performance improvement we observe mainly comes from the reduction in the number of cycles we run the job locally. Therefore, if we slightly reduce the effective speed of the servers, it should not significantly change the observed overall latency, meaning that many tasks still can be batched without needing extra cloud iterations.

\subsection{Other Features}

    This approach also has a few other advantages. First, based on the ratios between tasks falling into different iteration categories, we can get an estimation of how to allocate GPU resources. The workload of all the tasks in a certain group can be written as follows.
    \[w_{group} = n_{task} * n_{group}\]
    $n_{task}$ represents the number of jobs in this group and $n_{group}$ is the number of iterations we need to perform for every task in this group. The ratio between $w_{group}$ should serve as a good indicator for resource management since total work should be proportional to the number of jobs and work per job. Furthermore, we can free some GPU resources when the $\sum^{n_{groups}}_{n=0}{w_{i}}$ drops below a threshold.

\section{Evaluation}

\subsection{Experiment Setup}

\subsubsection{RegNet}
    
    For our experiments on RegNet, we used the pre-trained weights from PyTorch. It has 644.8 million weights, and a forward pass on it takes 374.57 GFLOPS \cite{PretrainedModels}. It achieved an accuracy of 88.228\% on ImageNet top-1 and 98.682\% on top 5. 

\subsubsection{Stable Diffusion Model}
    
    We utilized the PyTorch stable diffusion implementation from a GitHub repository\cite{SDModel}. Compared to the official version, this model is much more readable, making it easier for us to make changes. The PyTorch framework is also compatible with a range of platforms, which allows us to test with both professional and mobile GPUs with minimal code modifications. In the codebase, the model is conveniently divided into encoding, diffusion, and decoding phases. The diffusion iterations are performed in a loop. Thus, we can take the latent and context out from the corresponding step, serialize it, and share it with the client. We will show that the data transmitted is sufficient for reconstructing the image.

\subsection{Unit Tests}

\subsubsection{Transmission Cost}

    We perform transmission tests on both local and public networks. The public network results are obtained using a Google Cloud virtual machine located in Iowa, which is reasonably far from Chicago. We find for smaller tensors, the local network is faster due to the lower round-trip time. However, as the size increases, Google Cloud starts to outperform the local network. This is likely because of the higher-quality hardware Google uses for sending and receiving network packages.

\begin{figure}[h]
    \centering
    \includegraphics[width=8.5cm]{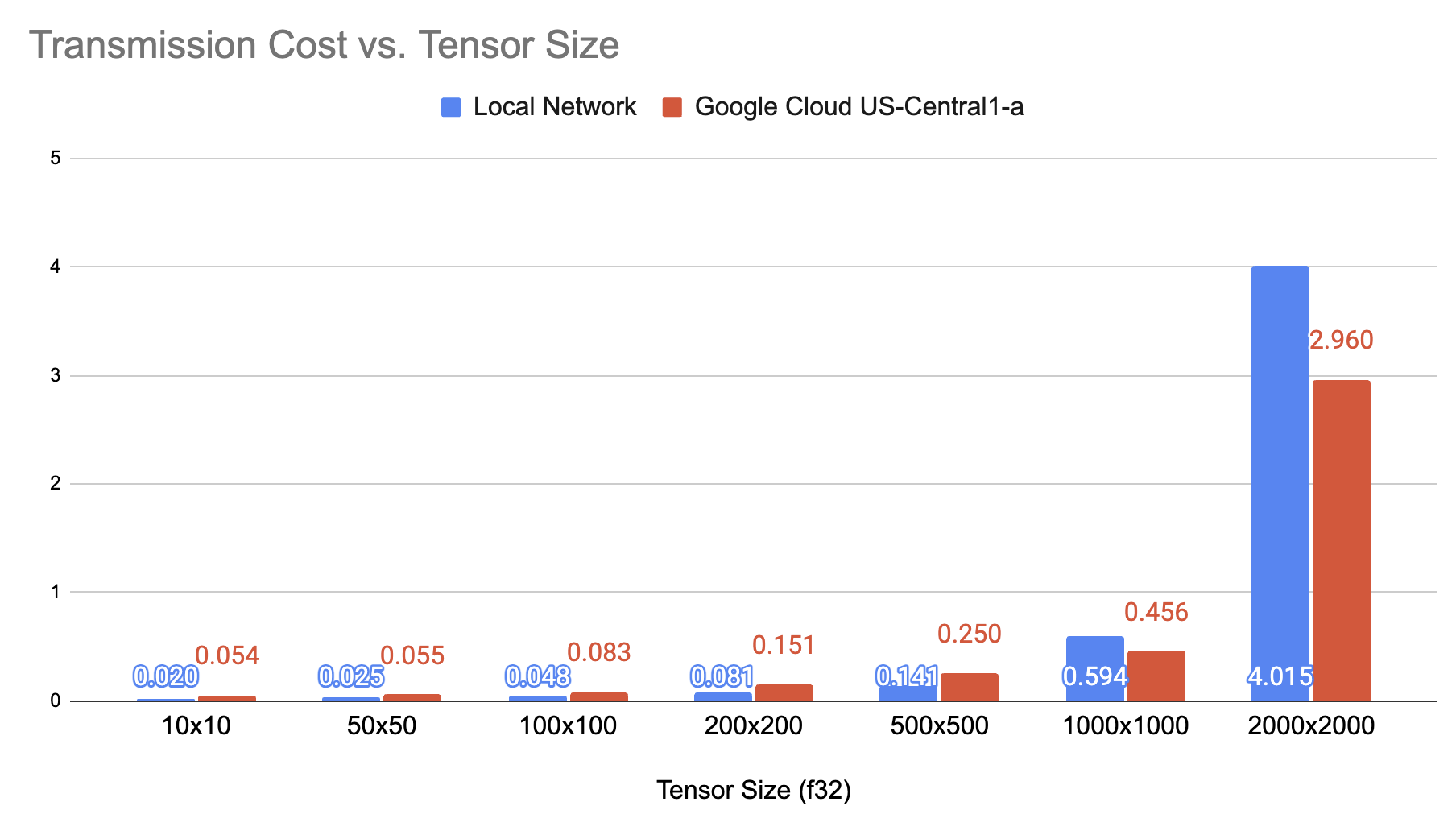}
    \caption{Transmission Cost vs. Tensor Size}
    \label{fig: transmission}
\end{figure}

    Another noticeable trend is that, while the costs increase slowly from 10x10 to 500x500 as shown in Fig \ref{fig: transmission}, it quickly grows afterward. A plausible explanation is that as the number of packets rises, the likelihood of failures increases, requiring expensive retransmission. Since the tensor sizes we are dealing with are smaller than 500x500, we believe it should not pose significant challenges for us. However, this test does present the importance of managing shared data for ML services.

\subsubsection{Serialization and Deserialization Cost}

\begin{figure}[h]
    \centering
    \includegraphics[width=8.5cm]{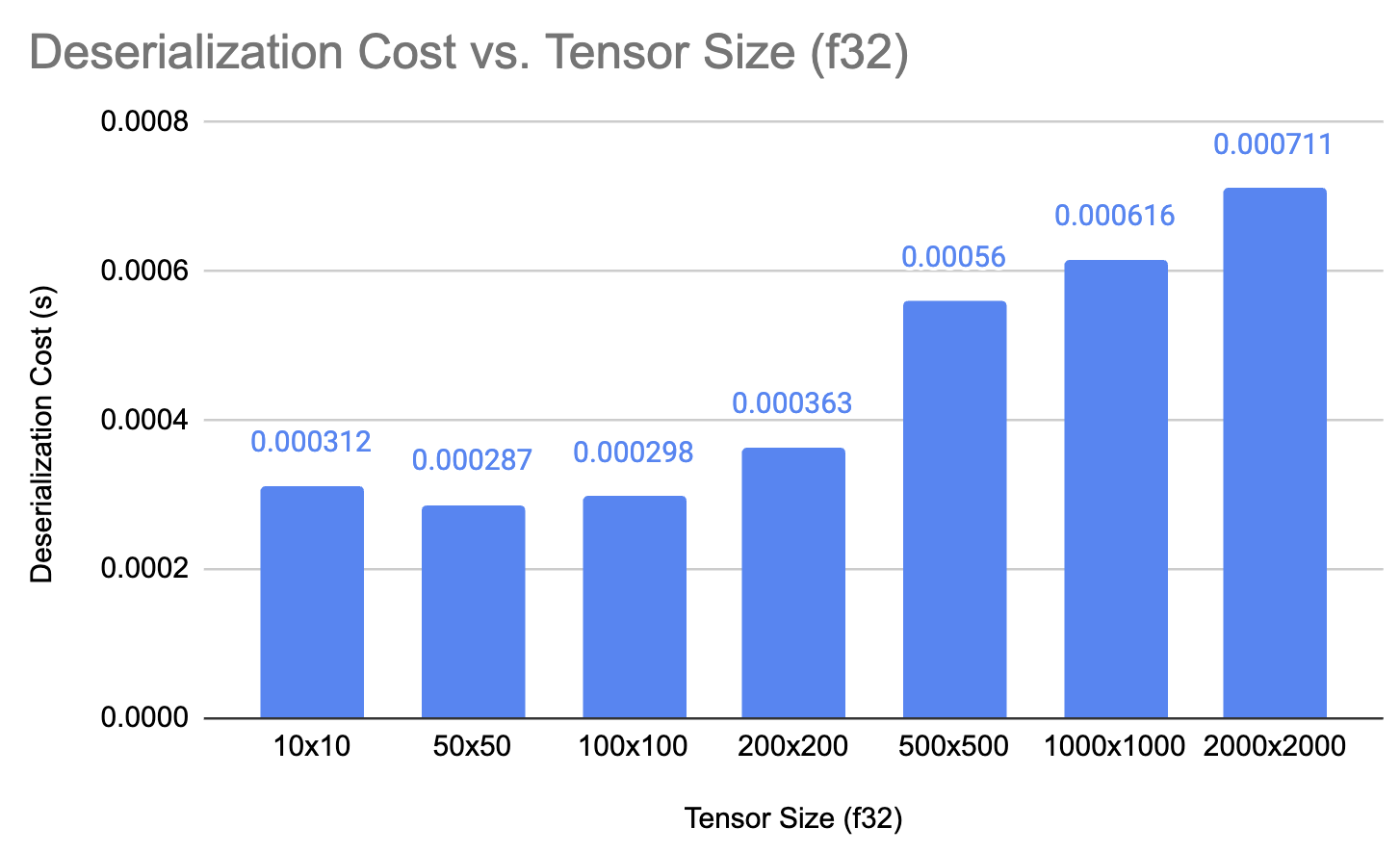}
    \caption{Deserialization Cost vs. Tensor Size}
    \label{fig: serialization}
\end{figure}

Based on our experiment results presented in Fig. \ref{fig: serialization}, the serialization cost for PyTorch tensors is almost negligible. This is most likely because the library simply points the buffer to the raw data. For the deserialization cost, we can see that it is not sensitive to the size of the tensor for the range that we are interested in, with the startup cost dominating from 10x10 to 200x200. However, it is clear that overall, the costs are insignificant when compared to the transmission cost.

\subsubsection{RegNet Performance}

\begin{figure*}[h]
    \centering
    \includegraphics[width=18cm]{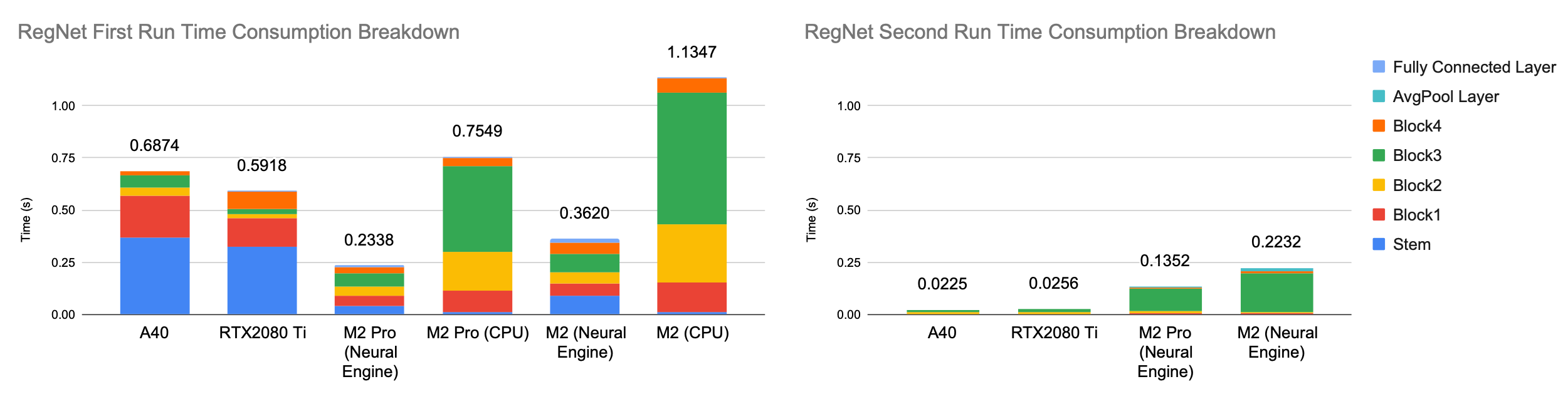}
    \caption{RegNet Performance on First and Second Inference}
    \label{fig: regnettime}
\end{figure*}

In the case of RegNet, we see that the results obtained in prior works start to break down\cite {Neurosurgeon}. As shown by Fig. \ref{fig: regnettime}, although professional GPUs are still significantly faster than mobile accelerators, the difference is not pronounced. Considering the fact that the round trip latency can easily exceed 100ms, it is hard to justify offloading in this case, especially given that RegNet is one of the most computationally demanding models for computer vision. Thus, we decided not to implement an offloading system for RegNet in this work.

Another interesting finding we have is that the startup cost seems to be larger on traditional GPUs. Fig. \ref{fig: regnettime} shows that mobile accelerators outperform GPUs on the first run. The increased time consumption mainly comes from the first two steps in the layer (stem and block 1). It is possible that these costs come from data transfer between CPU and GPU. Thus, as the model size grows, we need to reconsider the memory system design and data movement for machine learning.

\subsubsection{Stable Diffusion Performance}
\begin{figure*}[h]
    \centering
    \includegraphics[width=18cm]{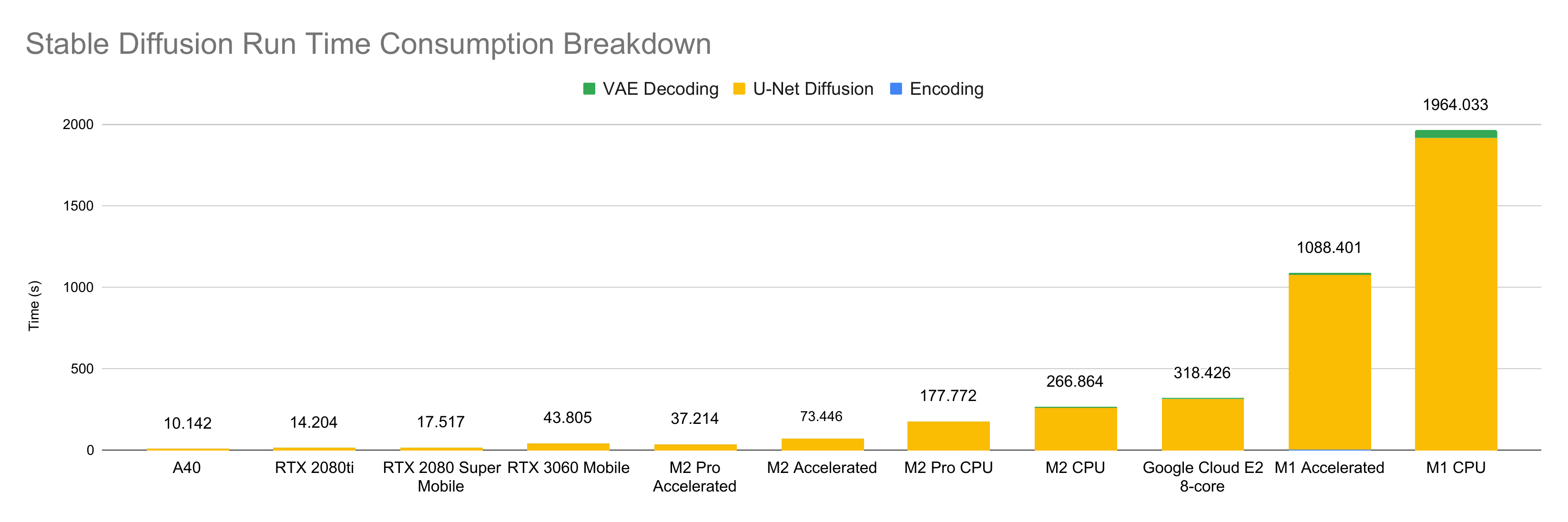}
    \caption{Stable Diffusion Performance}
    \label{fig: stabledifftime}
\end{figure*}

    We plot the stable diffusion performance we measured on a range of devices in Fig. \ref{fig: stabledifftime}. The resolution of the output images is 512x512, and the number of iterations we run is 50. For Apple silicon devices, we take advantage of their Metal Performance Shaders through PyTorch. We make a few important observations. First, compared to M1, M2 significantly reduces the time consumption. This result is consistent with the generational improvements many processor manufacturers are claiming. Therefore, ML service providers need to adapt to these changes quickly. Second, despite that, we still see huge latency reductions when we use professional GPUs to perform these jobs. We believe that is unlikely to change any time soon due to constraints like power consumption on mobile devices. Thus, accelerating generative models on the cloud will be fruitful. Third, devices without accelerators perform noticeably worse than those with accelerators. If the cloud treats all devices uniformly, it will waste a lot of GPU cycles. Lastly, the fact that the U-Net time makes up for most of the time consumption justifies our decision to divide the work at that step and ignore the encoding time on the cloud side.

\subsubsection{Effect of Batching}

\begin{figure}[h]
    \centering
    \includegraphics[width=8.5cm]{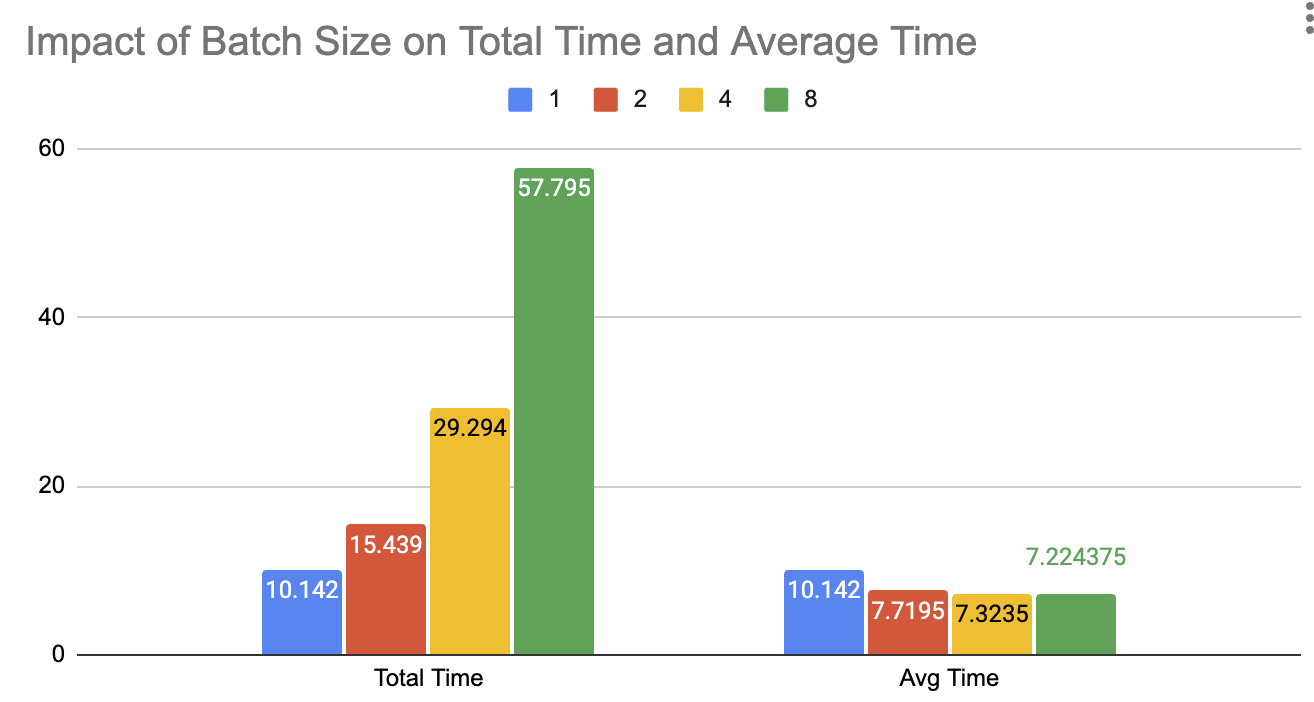}
    \caption{Impact of Batch Size of Total and Average Time}
    \label{fig: batch}
\end{figure}

    The traditional wisdom about batching tells us that it will increase resource efficiency because the total time increases slower than the number of inputs. However, in the setting we are considering, merging requests into a batch and launching them together can have an adverse impact: the response time for all the images in a batch will be determined by the batch completion time. Therefore, we need to evaluate the impact of batching carefully. Fig. \ref{fig: batch} shows the total time consumption and time consumption per image running stable diffusion on an NVIDIA A40 GPU. It is clear that we see a big reduction in average time consumption going from a batch size of one to two. However, further increasing it leads to diminishing returns. We will use that result in later sections to study the profitability of batching in our system.
    
\subsubsection{Preloading}
\begin{table}[h]
    \centering
    \begin{tabular}{l|c|c}
        \hline
        \hline
        Device & Time(s) & Diffusion Rate (iter/s)  \\
        \hline
        A40 & 10.142 & 4.930 \\
        \hline
        A40 Preloaded & 8.779 & 5.695 \\
        \hline
        RTX2080Ti & 14.204 & 3.520 \\
        \hline
        RTX2080Ti Preloaded & 11.793 & 4.240 \\
        \hline
        \hline
    \end{tabular}
    \caption{Time and Diffusion Rate w/wo Preloading}
    \label{tab: preloaded}
\end{table}

    For NVIDIA GPUs, we observe that they spend a non-negligible amount of time copying data from DRAM to VRAM. However, the server GPUs hosting these models can retain that data without having to load it every time they receive a request. The client-side can also overlap the loading time with some other operations it needs to perform. Therefore, it is important to measure the performance of those devices with the data preloaded. The results in Table \ref{tab: preloaded} show the benefits of adopting this technique.

\subsection{System Test}

\subsubsection{Image Quality}

\begin{figure}[h]
    \centering
    \includegraphics[width=8.5cm]{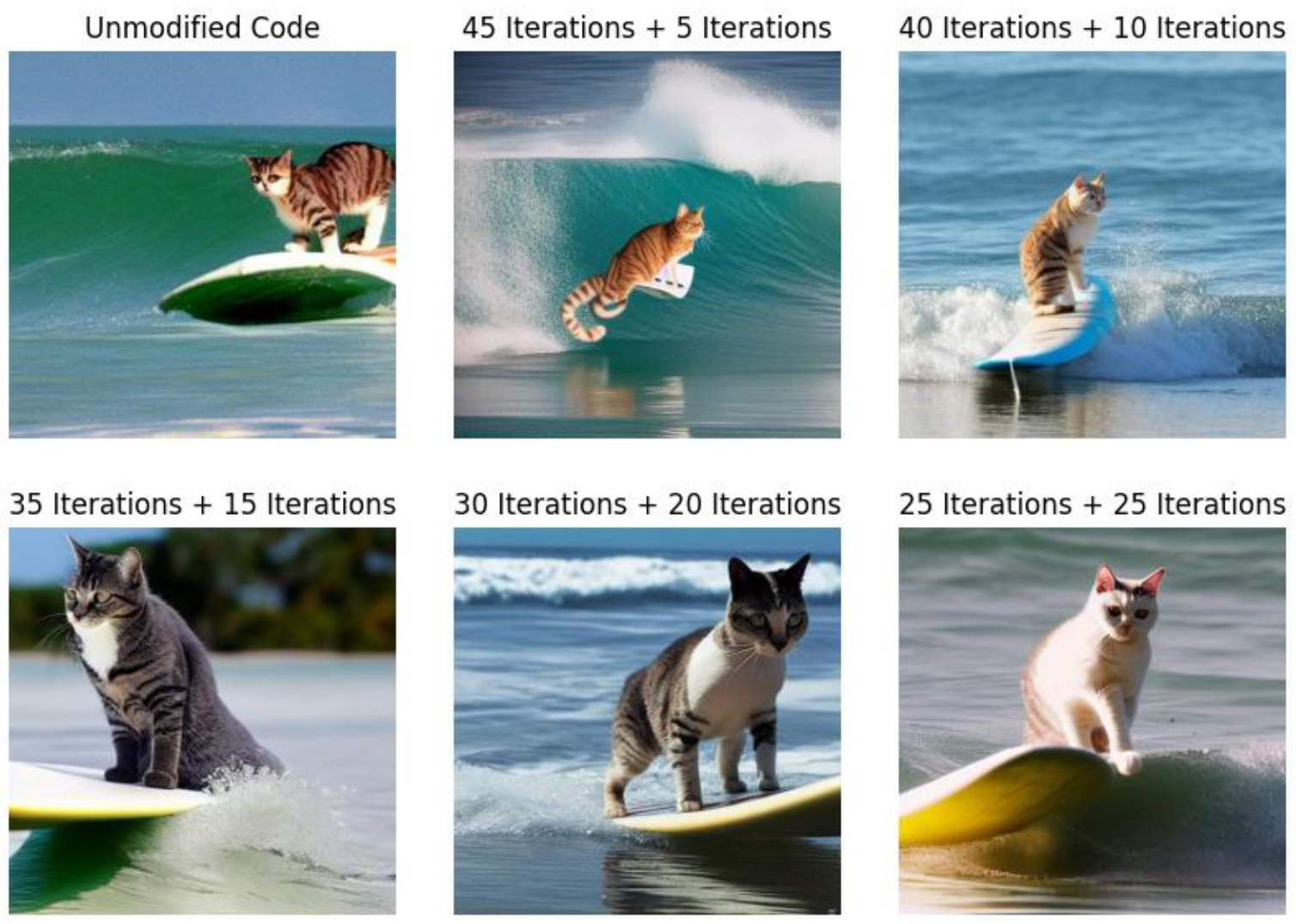}
    \caption{Quality of Generated Images under Various Settings}
    \label{fig: quality}
\end{figure}

    As shown by Fig. \ref{fig: quality}, separating the diffusion steps between the cloud and the local device does not make any noticeable impact on the quality of generated images. We should note that our code is based on an earlier release of the model. Therefore, the images are expected to have lower quality compared to the state-of-the-art.

\subsubsection{Time Consumption Reduction}

\begin{figure}[h]
    \centering
    \includegraphics[width=8cm]{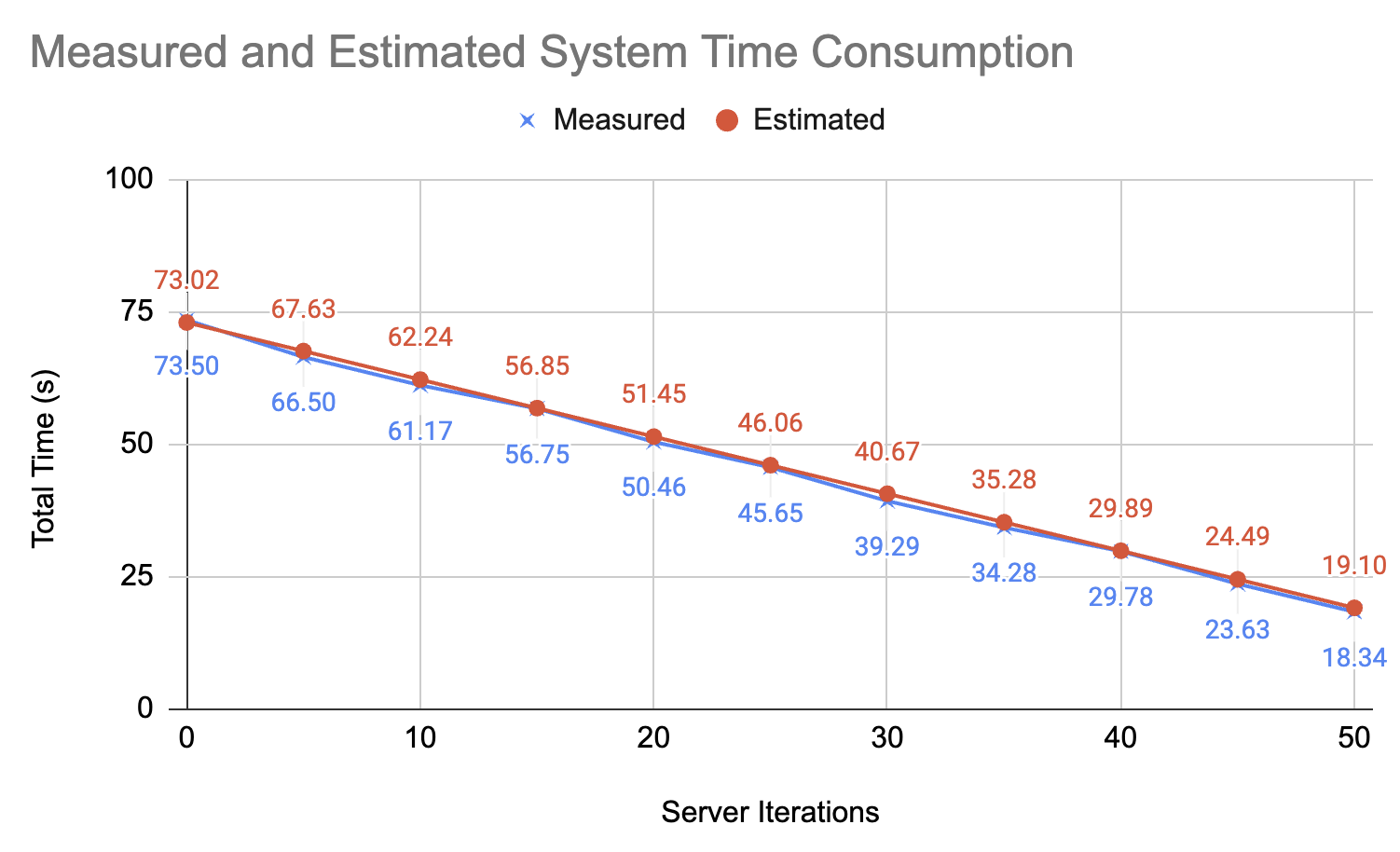}
    \caption{Measured Reduction in Time Consumption}
    \label{fig: time_reduce}
\end{figure}

    Fig. \ref{fig: time_reduce} shows the latency improvement for a MacBook Pro laptop with the M2 processor running stable diffusion when it is connected to another machine acting as the server with an RTX 2080 super mobile GPU. There are a few notable things. First, the blue line shows the measured system time, whereas the red line represents the estimated time derived from our cost model. The fact that those numbers align well suggests that our model is accurate. Second, since the RTX 2080 super graphics card takes around 17.5 seconds to run the complete network, we know that the overhead of our design is small since the system only uses 18.34 when we run all 50 iterations on the server. However, it should be noted that the results are obtained using a local network.

\subsection{Latency Enforcement}

\begin{figure}[h]
  \centering
    \includegraphics[width=8.5cm]{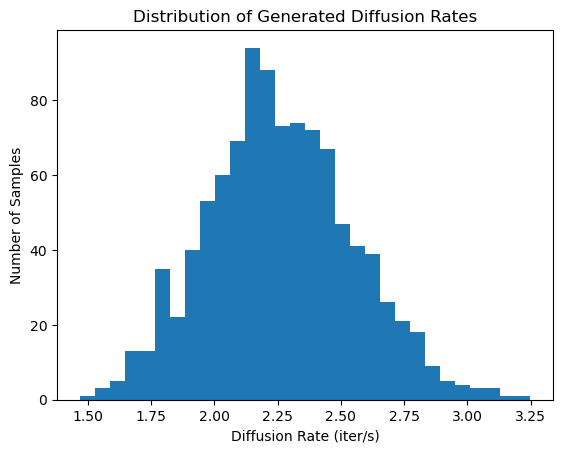}
    \caption{Distribution of Generated Diffusion Rates}
  \label{fig: diffrate}
\end{figure}

    We use the stable diffusion performance Apple presents to evaluate our proposed algorithm's ability to provide a consistent user experience. More specifically, we will generate requests based on the range of performance we observe on the v2-1 base model with 512x512 resolution. The slowest device they run the experiments on is the iPhone 12 Mini, which provides a diffusion speed of 1.44 iter/s. The fastest device they used is an M2 iPad Pro, delivering a throughput of 3.07 iter/s. Therefore, we generate a normal distribution of 1000 elements centered at 2.25 with a standard deviation of 0.28 (one-sixth of the range since three standard deviations cover 99.7\% of the samples). The histogram of generated samples is shown in Fig. \ref{fig: diffrate}.

    The server GPU performance estimate is based on a stable diffusion benchmark performed with the latest GPUs\cite{StableDiffusionPerf}. It shows the RTX4090 GPU can produce 75 images per minute when running 50 iterations for every image. This translates to a throughput of 62.5 iterations per second. Given that many AI-focused data centers use state-of-the-art GPUs, we believe it is reasonable to perform our experiments using that number.

    We assume a network latency of 300ms for transferring the tensors and prompt. The system will perform 50 iterations to guarantee the quality of output. The decoding cost for each sample is scaled appropriately based on its diffusion speed.

\begin{figure}[h]
  \centering
    \includegraphics[width=8.5cm]{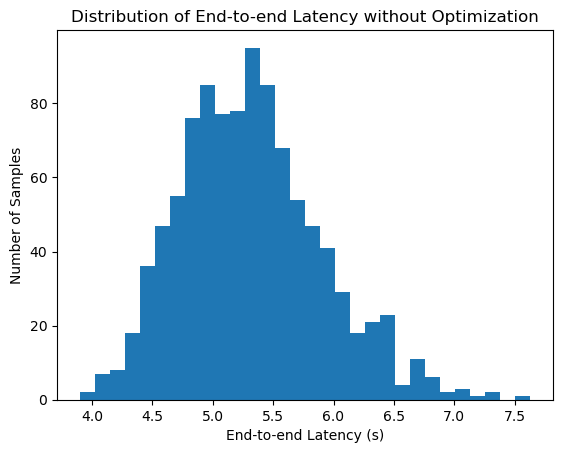}
    \caption{Distribution of Latency without Optimization}
  \label{fig: latnoopt}
\end{figure}

\begin{figure}[h]
  \centering
    \includegraphics[width=8.5cm]{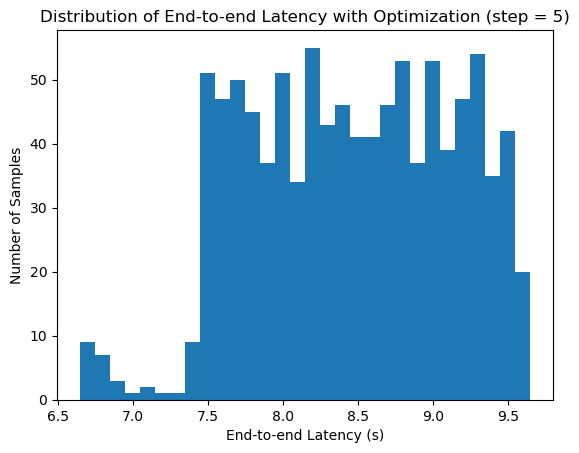}
    \caption{Distribution of Latency with Optimization}
  \label{fig: latwithopt}
\end{figure}

\subsection{Intelligent Batching}
\begin{figure}[h]
  \centering
    \includegraphics[width=8.5cm]{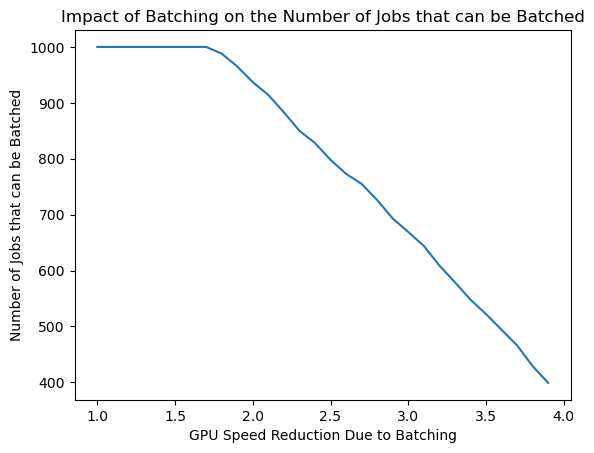}
    \caption{Impact of Batching Cost}
  \label{fig: batchcost}
\end{figure}

\begin{figure}[h]
  \centering
    \includegraphics[width=8cm]{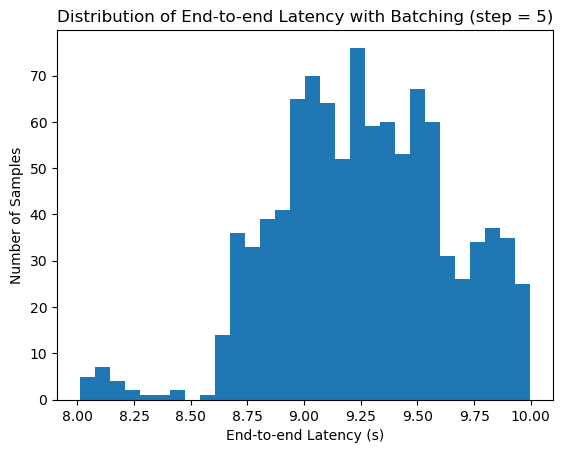}
    \caption{Latency Distribution after Batching}
  \label{fig: batch_lat}
\end{figure}

    As shown by Fig. \ref{fig: batchcost}, the number of examples that can run under the latency required by the SLA remains roughly constant until the batch cost exceeds 2.0. When the cost reaches 3.0, 60\% of all the tasks can still be batched together. This confirms our assumptions earlier that the benefits of cloud participation mainly come from the reduction in the number of cycles that need to run locally. Based on our previous measurements, the average time consumption only reduced significantly going from no batching to a batch size of 2. Therefore, we can easily adopt that setting, which leads to a roughly 1.6x increase in GPU processing time. Under that scenario, all the tasks can be batched without increasing the iteration count, meaning we are gaining free performance from this design. 

    Fig. \ref{fig: batch_lat} shows the overall latency distribution after we apply batching. Compared to Fig. \ref{fig: latwithopt}, we further cut the slack caused by the granularity of iteration counts. Since all the images can be batched together, we can expect a 30\% increase in efficiency based on the result from the NVIDIA A40 GPU, which experiences a drop in average processing time from around 10.0s to 7.5s. Table \ref{tab: result} shows the cloud GPU time needed for each scheduler.

\begin{table}[h]
    \centering
    \begin{tabular}{l|c}
        \hline
        \hline
        Scheduler & Cloud GPU time (s)\\
        \hline
        All work on Cloud & 800\\
        \hline
        Constant Iteration & 720\\
        \hline
        Variable Iteration & 600.96\\
        \hline
        Variable Iteration + Intelligent Batching & 487.06\\
        \hline
        \hline
    \end{tabular}
    \caption{Cloud GPU Time Under Various Scheduling Choices}
    \label{tab: result}
\end{table}

\subsection{Projection}
\begin{figure}[h]
  \centering
    \includegraphics[width=8.5cm]{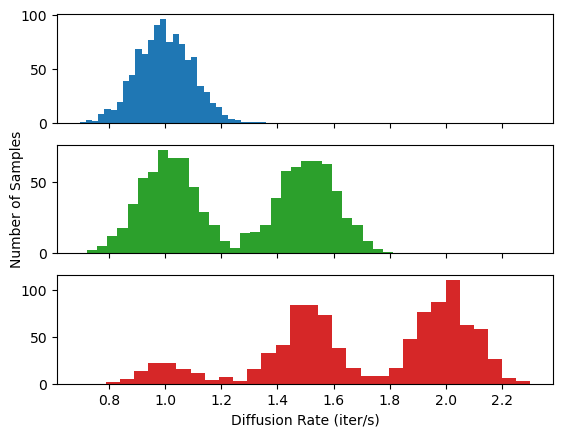}
    \caption{Projected Distribution of Generated Diffusion Rates}
  \label{fig: proj_diffrate}
\end{figure}

\begin{figure}[h]
  \centering
    \includegraphics[width=8.5cm]{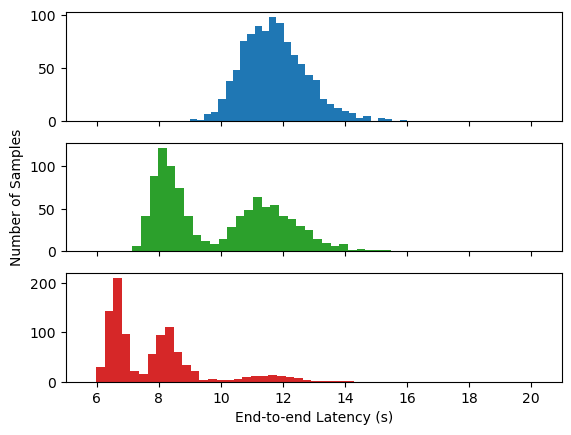}
    \caption{Projected Distribution of Latency without Optimization}
  \label{fig: proj_latnoopt}
\end{figure}

\begin{figure}[h]
  \centering
    \includegraphics[width=8.5cm]{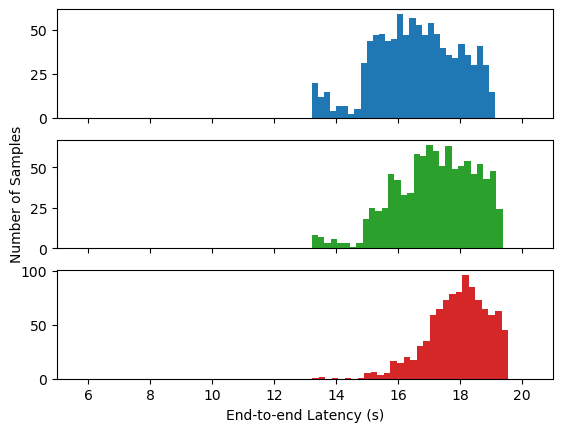}
    \caption{Projected Distribution of Latency with Optimization}
  \label{fig: proj_latwithopt}
\end{figure}

\begin{figure}[h]
  \centering
    \includegraphics[width=8.5cm]{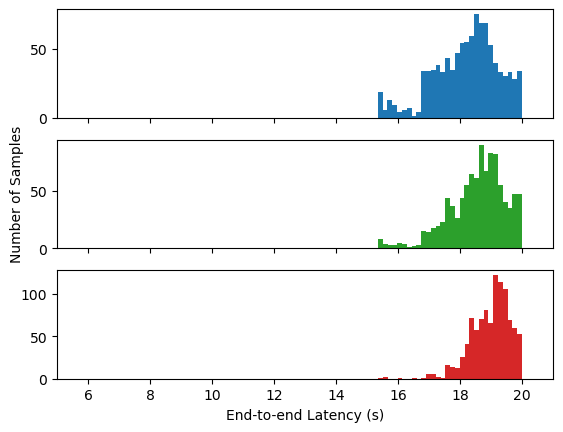}
    \caption{Projected Impact of Batching Cost}
  \label{fig: proj_batchcost}
\end{figure}

\begin{figure}[h]
  \centering
    \includegraphics[width=8.5cm]{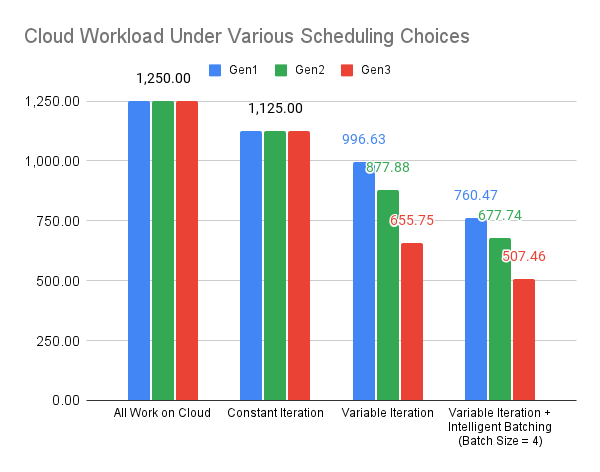}
    \caption{Projected Cloud GPU Time}
  \label{fig: proj_gputime}
\end{figure}

    To project the future when the model size further increases and the inference time increases on all devices. We set the mean diffusion rate of edge devices to 1.0 and the standard deviation to 0.1, and we set the estimated diffusion rate of a cloud GPU to 40 iterations per second. We also relaxed the time requirement to 20 seconds. To illustrate the increasing size of the vectors to transfer from cloud to edge devices, we increased the network transfer time to 500 milliseconds. We run the same scheduler described in the previous section with one thousand simulated devices. The projected distribution of generated diffusion rates, projected latency without optimization, with optimization, with batching, and the projected GPU times are presented in the blue bars in Figures \ref{fig: proj_diffrate}, \ref{fig: proj_latnoopt}, \ref{fig: proj_latwithopt}, \ref{fig: proj_batchcost}, \ref{fig: proj_gputime}. The projected cloud GPU time is reduced to 80\% and 61\% with variable iteration and variable iteration plus intelligent batching, respectively.

    We assume that a newer edge device with a 1.5 mean diffusion rate will be launched, and 50\% of the users will upgrade their devices. The distribution of the edge device will become the green bars in Figure \ref{fig: proj_diffrate}. With the same schedulers applied, we get the green bars in Figures \ref{fig: proj_latnoopt}, \ref{fig: proj_latwithopt}, \ref{fig: proj_batchcost}, and \ref{fig: proj_gputime}. Since the model has not changed, the cloud GPU time for running all work on the cloud has not changed. Further, there are still old devices used, and the cloud still needs to run 45 iterations if we run constant iterations on the cloud for all devices. However, our models can have benefits with more powerful edge devices joining, as the projected cloud GPU time is reduced to 70\% and 54\% with variable iteration and variable iteration plus intelligent batching, respectively.

    If 80\% of the 1.0 device users and 20\% of the 1.5 device users upgrade to a newer edge device that has a 2.0 mean diffusion rate, then the red bars in Figure \ref{fig: proj_diffrate}, \ref{fig: proj_latnoopt}, \ref{fig: proj_latwithopt}, \ref{fig: proj_batchcost}, and \ref{fig: proj_gputime} can represent the result. Similarly, the cloud GPU time for running all work on the cloud or running constant iterations on the cloud has not changed. Our models can have benefits with more powerful edge devices joining, as the projected cloud GPU time is reduced to 52\% and 41\% with variable iteration and variable iteration plus intelligent batching, respectively.

    This experiment shows that our scheduler, variable iteration w/wo intelligent batching, has the ability to increase the saving of cloud GPU time with the improvement of the inference ability of users' edge devices.

\section{Conclusion}

    Because of the rapid increase in model size, many machine-learning service providers have resorted to running their services in the cloud. However, we believe that decision fails to account for the improvement in the performance of mobile devices. Our results show that if we target the correct model and divide the work effectively, we can achieve low latency on a wide range of devices while significantly reducing the work that has to be performed on the cloud. We create a cost model for stable diffusion and design a scheduler that determines how many iterations need to run on the server based on the capability of the requesting device. We observe up to 60\% decrease in GPU demand compared to running everything on the server. We even reduce the size of the data that needs to be transferred if we only send the latents and perform decoding on the mobile device.

    We note that contrary to the findings of prior work\cite{Neurosurgeon}, not all models benefit from this offloading. For RegNet, one of the largest computer vision models, it runs sufficiently fast on mobile accelerators: the reduction in time consumption from using a state-of-the-art GPU can no longer make up for the time it takes to transmit the data. This result emphasizes the importance of insights into client devices when deploying machine learning inference services.

    Additionally, we propose a refinement called intelligent batching. By modifying the time model to account for the effect of batching, we allow tasks that can still satisfy the SLA after batching without increasing the cloud iteration to be grouped together. We show that because the latency improvement mainly comes from fewer cycles running on the client instead of the cloud's ability to run these tasks faster, a majority of tasks can be batched, leading to considerable efficiency improvement.

    Lastly, we make projections for future models and device performance, showing that our design responds well to variations. 

\section{Future Works}

    There are a few interesting directions worth exploring based on the findings of our work. First, the popularity of large machine learning models has driven an increase in the prevalence and performance of mobile accelerators. However, most users are not taking advantage of that hardware apart from when they run specific applications. Therefore, if we can achieve fog computing, sharing the idle capacity with older or power-constrained devices, we can expand their capabilities. There are a few challenges associated with such systems. The nodes, network connections, and resources are highly dynamic, making resource management difficult. There are also privacy and security concerns due to data sharing\cite{fog}. While our work proves that certain ML tasks benefit from offloading even after we account for the overhead, those issues still need to be resolved before this becomes a reality.

    Second, large model sizes impose higher requirements for the memory system. In order to avoid long run time, the parameters need to fit into the accelerator memory to avoid expensive swapping. To make matters even worse, we find that for NVIDIA GPUs, if the VRAM is smaller than the model, PyTorch rejects running the model on GPU. Therefore, cloud providers hosting multiple models must devise better memory management strategies. Mobile devices, on the other hand, have to adopt designs like unified memory to eliminate potential bottlenecks. Offloading decisions can also benefit from considerations about memory consumption since the performance degrades quickly when the model does not fit entirely into the memory. Occupying too much RAM can also adversely impact user experience.

    Third, there are some further refinements possible for the stable diffusion model. For example, if the user's prompt is shorter than the 77-token limit, we can transmit a smaller context vector since empty locations can be easily restored. Similarly, since generative models should fail gracefully with degrading input quality, it is possible that we can quantize the tensors we send or even continue the job when some packets are missing. If we can estimate the robustness of the model, we should be able to perform the transmission using inexpensive protocols like UDP. We can also create a policy that controls the SLA. When the usage is high or other higher-priority tasks arrive, we can relax the requirement to avoid dropping requests. When there is idle capacity, we put tighter bounds to improve latency.

\bibliographystyle{ACM-Reference-Format}
\bibliography{sample-base}


\end{document}